# SOME CONSTRAINTS ON GALAXY EVOLUTION IMPOSED

# BY THE SPECIFIC FREQUENCY OF GLOBULAR CLUSTERS

SIDNEY VAN DEN BERGH

Dominion Astrophysical Observatory

5071 West Saanich Road

Victoria, British Columbia

V8X 4M6, Canada

## ABSTRACT

Early-type dwarf galaxies are found to have high specific globular cluster frequencies which are similar to the high S values that are observed in luminous galaxies of types E and S0. It appears unlikely that cluster formation during mergers between dwarf disk systems can account for the high specific globular cluster frequency that is observed in early-type dwarf galaxies. Some alternative scenarios are briefly discussed.

## 1.  INTRODUCTION

Globular clusters are the oldest surviving sub-systems within galaxies. They therefore provide important information on the earliest phase of galaxy evolution. Particularly interesting constraints on galactic history are provided by the specific globular cluster frequency (Harris & van den Bergh 1981), which is defined as

$$S / N \times 10^{0.4 (M_V + 15)} , \qquad (1)$$

in which N is the total number of globulars and $M_V$ is the integrated visual magnitude of the parent galaxy of a cluster system. The present note will discuss the relations between S, the mean metallicity of globular cluster systems $< [Fe/H] >_{gc}$, and parent galaxy luminosity. A good review of previous work on this subject has been given by Harris (1991).

Almost 20 years ago, Toomre (1977) suggested that elliptical galaxies might have formed from merging spirals. It was subsequently pointed out (van den Bergh 1984, 1990) that the specific globular cluster frequency in giant elliptical galaxies is almost an order of magnitude higher than it is in giant spirals. If the number of globular clusters is conserved during mergers, then this observation

- 4 -might pose a serious problem for the merger scenario. This conundrum is somewhat mitigated by the fact that the young stellar populations in disk galaxies fade as they evolve. As a result, the specific globular cluster frequency, i.e. the number of clusters per unit luminosity, will increase with time. However, this effect is partly compensated for by the apparent brightening that takes place when the dusty gas in the late-type galaxies is consumed during star formation. Unfortunately, it is not possible to avoid the problems introduced by uncertain evolutionary and dust absorption effects by using the specific cluster frequency per unit baryonic mass. This is so because it is, in most galaxies, difficult to segregate baryonic mass from total mass.

Ten years ago, Schweizer (1986) suggested that the difference between the specific globular cluster frequencies in giant ellipticals and in giant spirals was due to globular cluster formation during collisions between spirals that were merging to produce ellipticals. Recently this idea has been explored in more detail by Ashman & Zepf (1992), Zepf & Ashman (1993) and others. Strong confirmation of the hypothesis that globular clusters can form when spirals collide appeared to have been provided by Hubble Space Telescope (HST) observations (Whitmore & Schweizer 1995) which showed large numbers of semi-stellar bright blue objects in the merging super giant disk galaxies NGC 4038/4039 (the Antennae). However, the interpretation of these objects as young globular clusters has recently been questioned by van den Bergh (1995a) who found that the blue objects in NGC 4038/4039 are (a) larger than typical Galactic globular clusters, and (b) have an open cluster, rather than a globular cluster, luminosity function. Taken at face value, these new results appear to suggest that collisions between disk galaxies result in a burst of <u>open</u> cluster formation. Clearly, it would be important to strengthen and confirm this conclusion with deeper HST images of NGC 4038/4039 and other colliding spirals.

Various objections to the idea that elliptical galaxies were formed by mergers of disk systems have been summarized by Ostriker (1980). Furthermore, Tremaine (1995) has recently emphasized that the apparent correlation between bulge luminosity and black hole mass, noted by Kormendy & Richstone (1995), is hard to understand in a scenarios in which E galaxies and bulges form from merging disk systems.

## 2. SPECIFIC CLUSTER FREQUENCY IN DWARFS

Most globular clusters have luminosities in the range $-10 < M_V < -5$. A complete census of globular cluster populations is therefore only possible in the nearest galaxies. Data on the specific globular cluster frequency S of Local Group members, which were kindly provided by Harris (1993), are listed in Table 1. The specific frequencies given in this Table exhibit a clear-cut dichotomy between disk galaxies, which have low values of S, and bulge-dominated elliptical/spheroidal galaxies, which (with the exception of M 32) show a much higher specific globular cluster frequency. Both giant and dwarf Local Group disk galaxies have $<S> \cdot 0.5$. A much higher value $<S> \cdot 6$ is found for the dwarf elliptical/dwarf spheroidal galaxies in the Local Group. The results discussed above suggest that <u>both dwarf and giant spheroidal systems have higher specific globular cluster frequencies than do disk galaxies of similar luminosities.</u> Some additional support for the conclusion that dE/dSph galaxies have high S values is provided by the recent discovery (Ibata, Gilmore & Irwin 1994) that the Sagittarius dwarf spheroidal galaxy appears to contain four globulars. Furthermore, the dwarf E1 galaxy NGC 3115B (Hanes & Harris 1986) also seems to exhibit a relatively high specific cluster frequency of $S = 6.3 \pm 3.1$.

The rather high S values that are observed in dE/dSph galaxies appear to militate against the suggestion by Fujimoto & Kumai (1991) that globular clusters form due to shock compression in collisions between clouds having velocity differences $> 100$ km s$^{-1}$. Specifically, such high-velocity collisions appear to be ruled out as a means of producing the five globular clusters in the Fornax dwarf, in which the internal velocity dispersion $F \sim 6$ km s$^{-1}$ (Seitzer & Frogel 1985).

## 3. COLLISIONS AND SPECIFIC FREQUENCIES

A strong constraint on models for the formation of giant elliptical galaxies is provided by spectroscopic observations (Worthy, Faber & Gonzalez 1992) which show that ["/ Fe] in ellipticals is larger than it is in spiral disks. The most plausible conclusion [see van den Bergh (1995c) for a complete discussion] from this observation is that ellipticals were assembled on a shorter time-scale than spirals. This is so because the gas in giant elliptical galaxies was polluted by "-elements produced by SNe II, but which condensed into stars before many Fe-producing SNe Ia had time to explode. The ["/ Fe] observations appear to rule out evolutionary scenarios in which the majority of giant elliptical galaxies formed from disk systems that had already transformed most of their gas into stars on a slow time-scale. It should, perhaps, be emphasized that such a picture in which E galaxies are assembled by merging mainly gaseous disks may not differ much from one in which ellipticals form from a single clumpy protogalaxy.

Sil'chenko (1993) has shown that ["/ Fe] is lower in disk galaxies than it is in typical ellipticals. Possibly, ellipticals exhibiting low ["/ Fe] are rare examples of early-type systems that did actually form from the late merger of disk galaxies that had already transformed most of their gas into stars.

Mergers between gaseous disk galaxies are expected to produce a burst of both star and cluster formation. However, as Harris (1994) has recently emphasized, such mergers will only increase S if the ratio of N (clusters) / N (young stars) is higher during star bursts than it is for quiescent galactic evolution.

The mean metallicity of globular clusters in individual giant elliptical galaxies (see Harris 1991 for a review) is lower than that of the bulk of the stars in their parent galaxies. At any galactocentric distance, most globular clusters therefore formed before average stars. In E galaxies, the rate of cluster formation must therefore have peaked before the main burst of star formation.

## 4.     DWARF ELLIPTICALS AND SPHEROIDALS

Why is the specific globular cluster frequency S as high in dwarf ellipticals and spheroidals as it is in giant ellipticals? As has been pointed out in § 2, high-velocity collisions between gas clouds cannot be invoked to explain the formation of globulars in dwarf galaxies. Furthermore, it is often assumed that dSph galaxies are former dIr galaxies which lost their gas as a result of ram pressure stipping (Lin

& Faber 1983) or by sweeping produced by supernova driven winds (Dekel & Silk 1986). They would therefore be expected to exhibit the low S values that are characteristic of d Ir galaxies, rather than the high ones that are actually observed. A possible way out of this conundrum is provided by the hypothesis that the rate of cluster formation in dwarf galaxies peaked before the rate of star formation. In such a scenario a violent initial burst of cluster formation would generate a strong wind which swept most of the gas out of the dwarf system, leaving only a relatively small amount of interstellar material to form stars. On the other hand, dwarf galaxies that did not undergo a significant initial burst would be expected to have lost less gas, from which second generation stars formed. As a result such dwarfs would be expected to end up with many stars, and hence with a low S value. In other words, high S values in dwarfs might provide fossil evidence for a significant amount of gas loss. Since dwarf spirals have nuclei, but dwarf irregulars do not (van den Bergh 1995d). Nucleated dwarfs with high S values (such as NGC 3115B) may once have been low-luminosity dwarf spirals that lost a significant amount of gas during their evolution. Van den Bergh (1993) has proposed that this gas loss in NGC 3115B was caused by the powerful radiation flux emitted by NGC 3115 (which contains a massive black hole) during its quasar phase.

## 5.     SPECIFIC CLUSTER FREQUENCY IN GIANTS

After excluding dwarf galaxies the data by Harris (1993) yield $<S> = 6.0$ for cD, E and S0 galaxies, compared to $<S> = 1.1$ for all spirals. The specific frequency $<S> = 9.3$ in nine cD/E galaxies is higher than that in 34 E + S0 galaxies for which $<S> = 5.1$. Furthermore, the specific frequency $<S> = 0.8$ in eight late-type spirals (Sb - Scd) is significantly lower than the value $<S> = 1.9$ in three bulge-dominated Sa and Sab galaxies. In summary, it appears that the specific globular cluster frequency in E and S0 galaxies, in which $<S> = 5.1$, is ~ 6 times greater than that in disk-dominated spirals which have $<S> = 0.8$. This difference will, however, be reduced to a factor of ~ 3 after fading due to stellar evolution (Bruzual & Charlot 1993) and brightening due to consumption of dusty gas in star formations are taken into account (van den Bergh 1995c). Ashman & Zepf (1992) and Zepf & Ashman (1993) have elaborated on the interesting suggestion by Schweizer (1986) that the difference between the S values in luminous ellipticals and spiral galaxies is due to cluster formation during the collisions between spirals that resulted in the mergers from which ellipticals were born. A possible check on this suggestion is provided by observations of the metallicity distribution among globular clusters associated with elliptical galaxies. In the Ashman & Zepf scenario, the primordial globular clusters should mostly be metal-poor, whereas those formed later form the gas in evolved colliding spirals should be metal-rich. Since (after correction for stellar evolution and dust removal) S (ellipticals) • 3 S (spirals), one would expect the population of metal-rich globulars in ellipticals to be (on average) twice as large as that of the metal-poor primordial ones. What little information is presently available suggests that metal-rich and metal-poor clusters actually occur with roughly equal frequencies in E galaxies. Ashman & Zepf (1993) find that NGC 4472 contains roughly equal populations of metal-poor clusters peaking at [Fe/H] = -1.7 and metal-rich ones with a maximum frequency at [Fe/H] = -0.5. In NGC 5128, 33 (55%) of clusters belong to a blue population with [Fe/H] ~ -1.2, and 27 (45%) to a red population with [Fe/H] ~ -0.1 (Ashman, Bird & Zepf 1994). In NGC 3923 (Zepf, Ashman & Geisler 1995), 54% of the 143 globulars have [Fe/H] ~ -0.94 and 46% are metal-rich with [Fe/H] ~0.00. Finally, in M 87 = NGC 4486 (Elson & Santiago 1995) the situation is more complex. Among the brightest globulars with V < 23.5, 72% are metal-poor with [Fe/H] peaking at -1.5, and 28% are members of a metal-rich population with a maximum frequency at [Fe/H] ~ -0.5. However, among the fainter clusters, the situation is reversed with 42% belonging to the metal-poor group and 58% to the metal-rich group. In summary, it appears that presently available data on the frequency distribution of metallicities among globular clusters in elliptical galaxies show a slightly smaller frequency of metal-rich clusters than might have been expected from the merging spirals model. However, three-peaked frequency distributions, such as that found by Ostrov, Geisler & Forte (1993) in NGC 1399, suggest that the evolutionary history of some giant elliptical galaxies may have been quite complex.

## 6.    CLUSTER METALLICITIES

Presently available data on cluster systems, for which both S and $<$ [Fe/H] $>_{gc}$ are reasonably well determined, are listed in Table 2. The entries in this table were mostly drawn from a recent compilation by Perelmuter (1995), which includes important new information on the M 81 cluster system obtained by Perelmuter, Brodie & Huchra (1995). A plot of the mean metallicity of cluster systems versus the total number N of globular clusters in each system is shown in Fig. 1. This Figure shows that the mean metallicity of cluster systems rises as the number of clusters in the system increases. The correlation coefficient between log N and $<$ [Fe/H] $>_{gc}$ is found to be $r = 0.88 \pm 0.05$. For the data in Table 2 the correlation coefficient between parent galaxy luminosity $M_V$ and $<$ [Fe/H] $>_{gc}$ is $r = 0.83 \pm 0.08$. This shows that the mean metallicity of globular cluster systems correlates at least as well with the total amount of cluster formation as it does with the total amount of star formation in the parent galaxy of the system.

Kumai, Hashi & Fujimoto (1993) found a correlation between the specific globular cluster frequency S and the local galaxy density $D_g$ (defined as the space number density of galaxies around the galaxy under consideration). This apparent correlation might be due primarily to the fact that (1) E galaxies exhibit high S values, and (2) that ellipticals occur preferentially in regions of high $D_g$. However, the observation that elliptical galaxies in the Virgo cluster tend to have higher S values than do E galaxies in other environments (Harris 1991) might also contribute to the effect noted by Kumai et al.

7.      **METALLICITIES OF GLOBULAR CLUSTERS**

Inspection of Table 1 shows that $<[Fe/H]>_{gc}$, the <u>mean</u> metallicity of globular cluster systems, exhibits a wide range (-2.05 to -0.31) in early-type galaxies. Late-type systems show a much smaller range (-1.66 to -1.21) for this parameter. This difference is mostly, but probably not entirely, due to the fact that giant ellipticals contain more very metal-rich clusters than do less bulge-dominated galaxies. Figure 2 shows a plot of [Fe/H] versus the perigalactic distance $R_p$ (van den Bergh 1995b) for individual Galactic globular clusters. This figure strikingly illustrates the dependence of cluster metallicity on perigalactic distance. At any value of $R_p$, the metallicity of Galactic globular clusters is seen to fall below the line

$$[Fe/H] = -0.7 \log R_p - 0.27, \qquad (2)$$

in which $R_p$ is in kpc. A monotonic relation of this type would be expected if the metallicity of the gas in a galaxy depends on escape velocity, i.e. on the depth of the potential well in which it is located (Franx & Illingworth 1990). On such a picture, the observation that the highest values of $<[Fe/H]>_{gc}$ occur in elliptical galaxies would be accounted for by the fact that these objects have the largest masses, and hence the deepest potential wells. Alternatively, it might be assumed that E galaxies formed from the mergers of the most massive ancestral objects.

## 8. CONCLUSIONS

Both giant elliptical galaxies and dE and d Sph galaxies are observed to have high specific globular cluster frequencies. This contrasts with the situation in both giant and dwarf late-type disk systems in which S is found to be low. The high frequency of globulars in both giant and dwarf early-type galaxies might be due to either (1) a high initial specific globular cluster frequency in all early-type non-disk galaxies, or (2) to separate causes in early-type giants and in early-type dwarfs. A high value of S in dE and d Sph galaxies might, for example, be due to preferential formation of massive clusters during an early burst of star formation. On the other hand, ejection of interstellar gas by supernova-driven stellar winds might reduce the total amount of late-time star formation, and therefore increase S. If this hypothesis is correct, then one might attribute the low S values observed in dwarf disk galaxies to the "dilution" of S produced by later generations of star formation in a quiescent environment that favored star formation over production of massive globular clusters.

The suggestion that the high S values, which are currently observed in most giant ellipticals, is due to cluster formation during collisions and subsequent mergers of giant spiral galaxies meets with four possible difficulties: (1) The luminosity function of the numerous clusters produced by the recent collision of the giant spirals NGC 4038/4039 has resulted in the formation of objects that appear to have an <u>open</u> cluster luminosity function. (2) The ratio of the number of metal-poor clusters to the number of metal-rich globulars in giant E galaxies is possibly larger than one would have expected from the collisions/mergers scenario. (3) The high ["/Fe] values that are observed in gE galaxies appear to rule out scenarios in which most giant ellipticals formed from mergers of spiral galaxies in which most of the gas had already been transformed into stars. The ["/Fe] observations are, however, consistent with a scenario in which gE galaxies formed from collisions between many gaseous disk galaxies, or from a single clumpy protogalaxy. Finally, (4) the merger scenario does not provide an obvious explanation for the apparent correlation between the masses of black holes and the luminosities of their parent bulges/galaxies.

I am indebted to Dr. Jean-Marc Perelmuter for useful discussions, to Keith Ashman, Michael West and Stephen Zepf for helpful exchanges of e-mail and to David Duncan for drawing the figures.

**TABLE 1 - Local Group Cluster Frequencies**

| Name | Type | S |
|---|---|---|
| Galaxy | Sbc | 0.5 ± 0.1 |
| M 31 | Sb | 0.7 ± 0.2 |
| M 33 | Sc | 0.6: |
| LMC | Ir/SBc | 0.5: |
| SMC | Ir | 0.4 ± 0.2 |
| Disk mean | | 0.5 |
| NGC 205 | E6p | 2.3 ± 0.3 |
| M 32 | E2 | < 0.8 |
| NGC 185 | dE0 | 6.5 ± 1.6 |
| NGC 147 | dE4 | 4.0 ± 1.0 |
| Fornax | Sph | 17.0 ± 7.0 |
| Elliptical mean | | 6.0 |

**TABLE 2 - Globular Cluster Systems**

| Galaxy | Type[a] | $M_V$[b] | $N_{gc}$[b] | S | $<[Fe/H]>_{gc}$[b] |
|---|---|---|---|---|---|
| Fornax | d Sph | -13.7 | 6 ± 1 | 19.9 ± 3.3 | -1.85 ± 0.1 |
| N 147 | d E5 | -15.0 | 4 ± 2 | 4.0 ± 2.0 | -2.05 ± 0.4 |
| N 185 | d E3p | -15.2 | 8 ± 2 | 6.6 ± 1.7 | -1.65 ± 0.2 |
| N 205 | S0/E5p | -16.5 | 9 ± 1 | 2.3 ± 0.3 | -1.45 ± 0.1 |
| SMC | Im | -16.9 | 2: ± 1: | 0.3 ± 0.2 | -1.40 ± 0.1 |
| LMC | SBm | -18.6 | 15: ± 5: | 0.5 ± 0.2 | -1.66 ± 0.1 |
| M 33 | Sc | -19.2 | 30: ± 10: | 0.6 ± 0.2 | -1.40 ± 0.2 |
| M 81 | Sb | -21.2 | 210 ± 30 | 0.7 ± 0.1 | -1.48 ± 0.2 |
| N 1399 | E1 | -21.2 | 3600 ± 1100 | 11.9 ± 3.6 | -0.85 ± 0.1 |
| Galaxy | Sbc | -21.3 | 160 ± 20 | 0.5 ± 0.1 | -1.35 ± 0.05 |
| M 31 | Sb | -21.7 | 270 ± 50 | 0.6 ± 0.1 | -1.21 ± 0.05 |
| N 5128 | S0 + Sp | -22.0 | 1700 ± 400 | 2.7 ± 0.6 | 0.80 ± 0.2 |
| N 3923 | E4/S0 | -22.1 | 4300 ± 2000 | 6.2 ± 0.9 | -0.56 ± 0.2 |
| N 4649 | S0 | -22.2 | 5800 ± 1300 | 7.6 ± 1.7 | -1.10 ± 0.2 |
| M 87 | E0 | -22.4 | 16000 ± 4000 | 17.5 ± 4.4 | -1.00 ± 0.2 |
| N 4472 | E1/S0 | -22.6 | 7400 ± 2000 | 6.7 ± 1.8 | -0.80 ± 0.3 |
| N 3311 | E0/cD | -22.8 | 13700 ± 5500 | 10.4 ± 4.1 | -0.31 ± 0.4 |

[a] Mainly from Sandage & Tammann (1981)

[b] From Perelmuter (1995)

# FIGURE LEGENDS

Fig. 1  Correlation between total number of globular clusters N and mean cluster metallicity. Symbols are as in Fig. 2. Disk systems are plotted as open squares, dwarf ellipticals as open circles and giant ellipticals as filled circles. The figure shows that cluster metallicity correlates with the total amount of cluster formation.

Fig. 2  Galactic globular cluster metallicity versus perigalactic distance. At a given perigalactic distance, globulars all have metallicities below the value given by Eqn. (3). Metal-rich clusters are seen to be absent at large perigalactic distances.

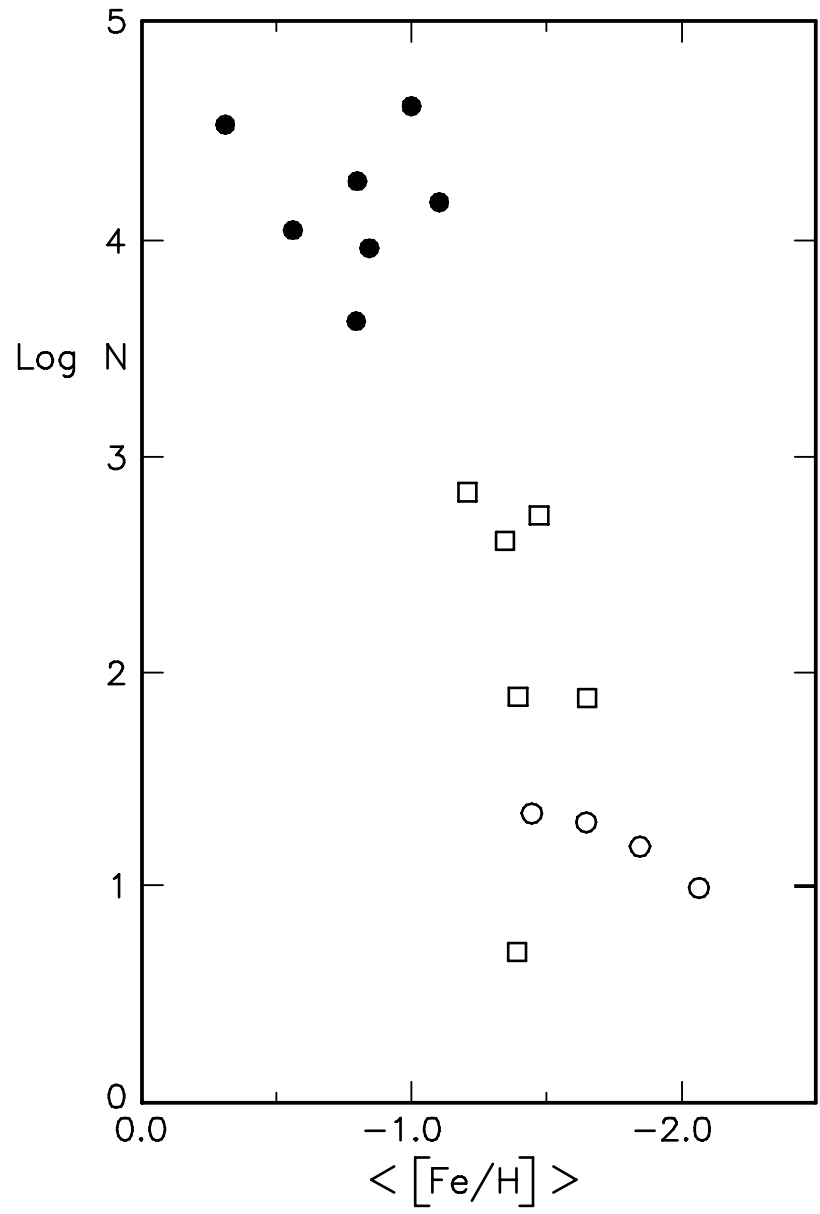

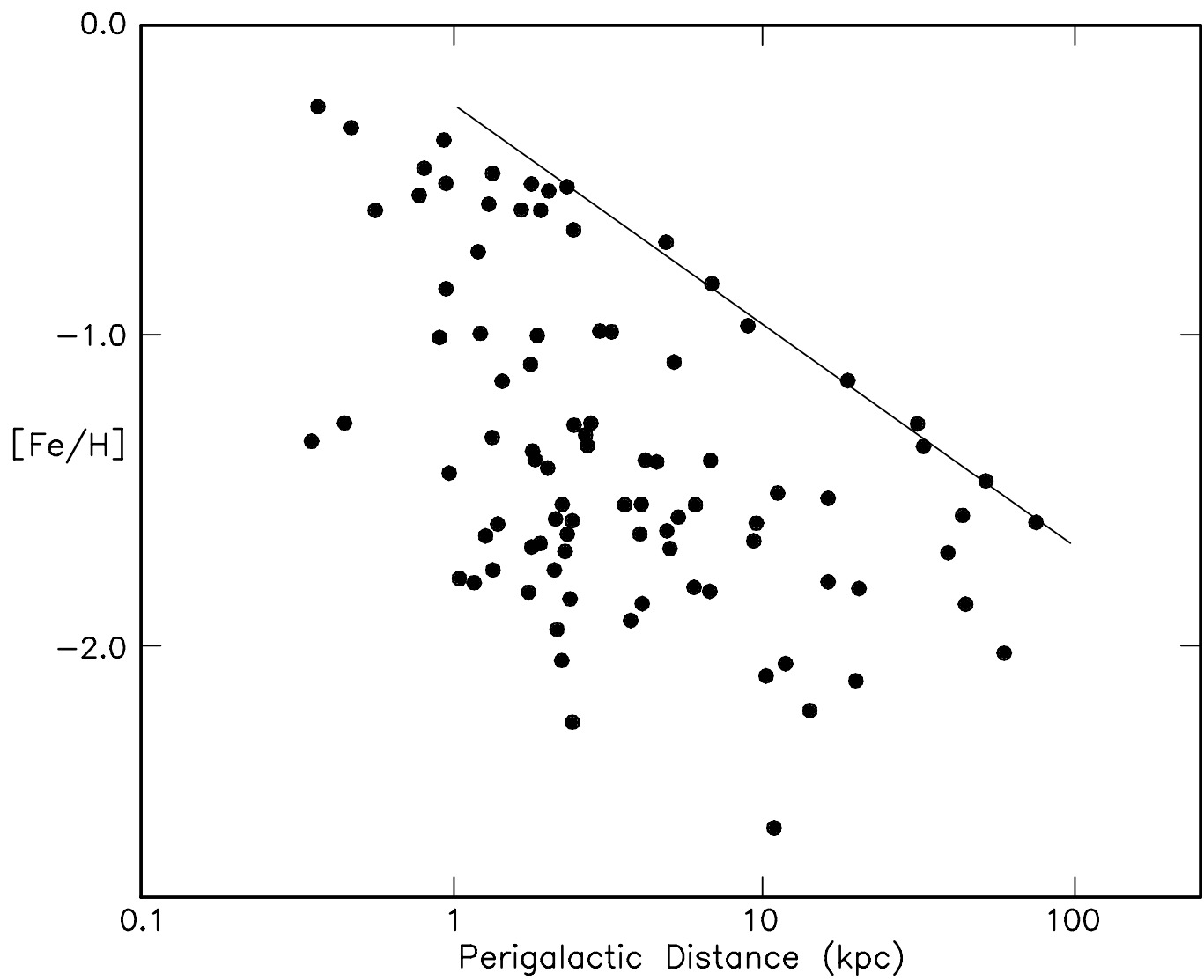